\documentclass[twocolumn,english,aps,prl,showpacs,superscriptaddress,longbibliography]{revtex4-1}
\usepackage[T1]{fontenc}
\usepackage[latin9]{inputenc}
\usepackage{color}
\usepackage{amsmath}
\usepackage{amssymb}
\usepackage{graphicx}
\usepackage{wasysym}
\usepackage{esint}
\usepackage{comment}
\makeatletter
\usepackage{babel}
\usepackage{dsfont}
\usepackage[dvipsnames]{xcolor}

\usepackage{hyperref}
\def\equationautorefname~#1\null{Equation (#1)\null}

\makeatother

\usepackage{babel}
\IfFileExists{lmodern.sty}{\usepackage{lmodern}}{}
\begin{document}

\title{Optically induced flat bands in twisted bilayer graphene}

\author{Or Katz}
\email[Corresponding author: ]{or.katz@weizmann.ac.il}
\affiliation{Physics Department, Technion, 3200003 Haifa, Israel}

\affiliation{Rafael Ltd, IL-31021 Haifa, Israel}

\affiliation{Department of Physics of Complex Systems, Weizmann Institute of Science, Rehovot 76100, Israel}

\author{Gil Refael}
\affiliation{Institute of Quantum Information and Matter and Department of Physics, California Institute of Technology, Pasadena, California 91125, USA}

\author{Netanel H. Lindner}
\affiliation{Physics Department, Technion, 3200003 Haifa, Israel}

\begin{abstract}
Twisted Bilayer Graphene at the magic twist angle features flat energy bands,  which lead to superconductivity and strong correlation physics. These unique properties are typically limited to a narrow range of twist angles around the magic angle with a small allowed tolerance. Here we report on a mechanism that enables flattening of the band-structure using coherent optical illumination, leading to emergence of flat isolated Floquet-Bloch bands. We show that the effect can be realized with relatively weak optical beams at the visible-infrared range (below the material bandwidth) and persist for a wide range of small twist angles, increasing the allowed twist tolerance by an order of magnitude. We discuss the conditions under which these bands exhibit a non-zero Chern number. These optically induced flat bands could potentially host strongly-correlated, non-equilibrium electronic states of matter.


\end{abstract}
\maketitle
Van-der-Waals heterostructures are a prominent tool for discovery of emergent phenomena in condensed-matter physics. These materials allow for a considerable degree of control in their physical structure, being formed by stacking of individual atomic layers \cite{geim_VDW_review,Novoselov2016,VDW_review_2016}. Stacking different atomic layers with a relative angular twist has become a salient mechanism in structuring the energy bands of these materials \cite{MacDonald_2019,Neto_2007,Dean2018,Kaxiras2017}. This twist forms a slowly varying moir\'e pattern, which modulates the inter-layer electronic potential. At certain twist angles, the bands near charge-neutrality point (CNP) can become flat and relatively isolated from other bands \cite{Neto_2012,MacDonald2011,Ashvin2019,Song2019}. These flat bands have recently attracted considerable attention, with the discovery of superconductivity, correlated insulating states and ferromagnetism, which emerge at low temperatures \cite{PJH_superconductivity,PJH_correlated_ins,Macdonald2017,Dean_2019,DGG_ferro_2018,Bernevig_phonons}.

Twisted bilayer graphene (TBG) exhibits isolated flat bands when twisted near the magic twist angle, $\theta_{\mathrm{m}}\approx1.1^{\circ}$. At smaller twist angles $\theta<\theta_{\mathrm{m}}$, a larger moir\'e pattern is formed, eliminating the energy gap to distant energy levels and increasing the bandwidth of the bands near CNP \cite{MacDonald2011,koshino2017}. Therefore, the discovery of strong correlated phenomena has been limited to a small range of twist angles near the magic angle, where the band-structure features narrow gaped bands.

Floquet-engineering with optical fields is a valuable technique that could induce topological band-structures and electronic correlations in various materials \cite{Floquet-1,Floquet-2,Floquet-3,Floquet-4,Floquet-5,Floquet-6,Floquet-7,Floquet-8,Floquet-9,Floquet-10,Floquet-11,Moore-2017,Fertig2011,Sentef2015, Klinovaja2016, Liu2018, Lindner2019}. Floquet engineering of TBG at large twist angles has been considered as a technique to control the topology of the bands near CNP  \cite{Floquet_graphene}. However, the idea of significantly reducing the bandwidth of certain bands in the band-structure, generating flat-bands, has never been realized in Van der Waals heterostructures with optical fields. While the mechanism of dynamic-localization could be theoreticaly applied  \cite{dynamic_localization}, it practically requires extremely high fields which render it unfeasible.

\begin{figure}[b]
\begin{centering}
\includegraphics[viewport=0bp 0bp 711bp 233bp,clip,width=8.6cm]{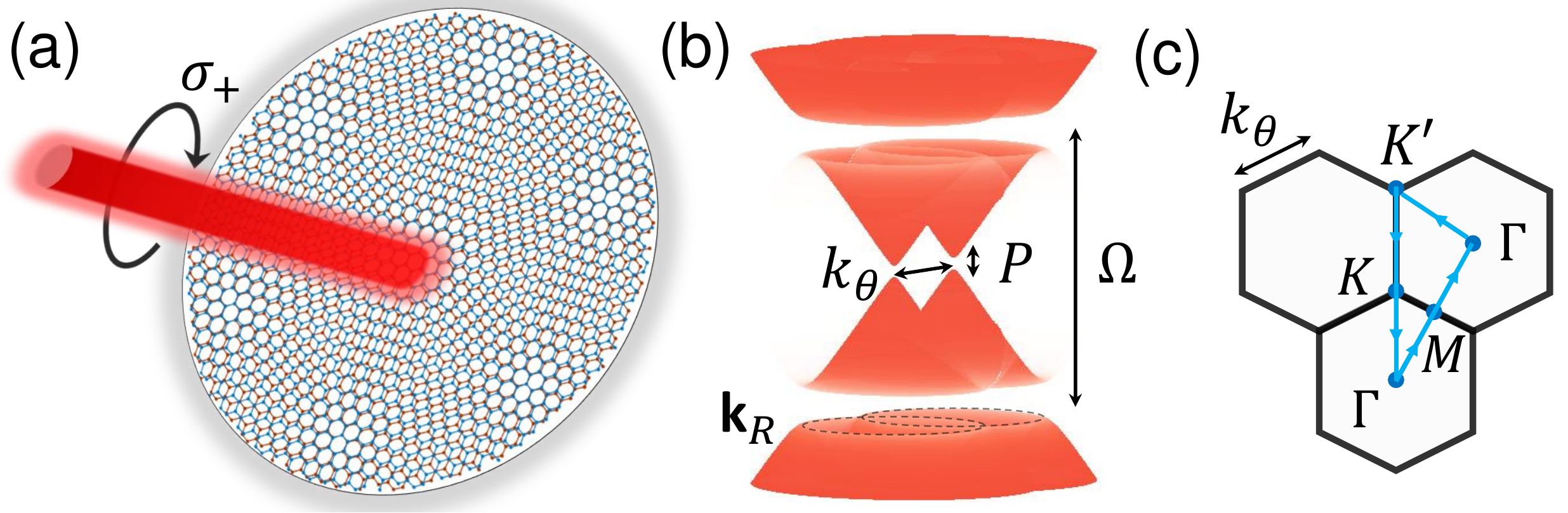}
\par\end{centering}
\centering{}\caption{Schematics of Floquet Twisted Bilayer Graphene. (a) Twisted Bilayer
graphene (TBG) driven by a circularly polarized optical laser field.
(b) Illustration of the band-structure of optically driven TBG with zero interlayer coupling near the $K$ points of the two layers. The laser frequency $\Omega$ in the visible range is much smaller than graphene bandwidth, opening a photo-induced gap $P$ at the $K$ points of the two layers while avoiding side-band transitions near these points. Nonzero interlayer hopping
leads to hybridization of the Dirac cones. (c) Moir\'e Brillouin Zone with the trajectory $K'\rightarrow K\rightarrow\Gamma\rightarrow M\rightarrow\Gamma\rightarrow K'$.
\label{fig: illustration} }
\end{figure}

Here we consider small-angle twisted bilayer graphene driven with optical fields, as shown in Fig.~\ref{fig: illustration}. We demonstrate that a driving laser could improve the flatness of the bands near the CNP, even at twist angles smaller than the magic angle. We further show that the driving field opens a gap between the emerging flat-bands, increases the gaps separating them from other bands, and could induce a nontrivial Berry curvature. The presented effect is found robust for lattice relaxation and could potentially be implemented with relatively weak sub-bandwidth optical fields in the visible range.

\begin{figure}[t]
\begin{centering}
\includegraphics[viewport=0bp 0bp 663bp 521bp,clip,width=8.6cm]{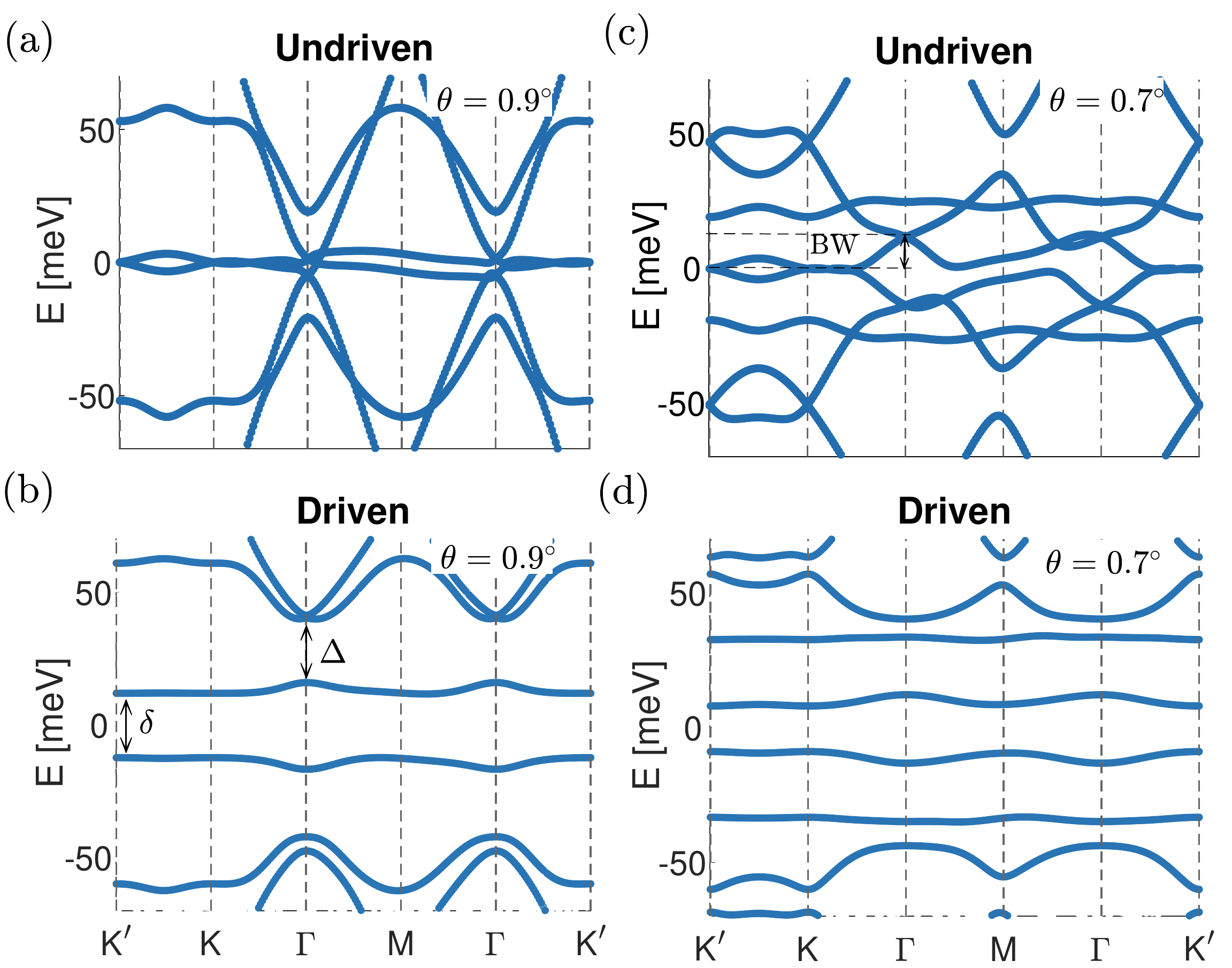}
\par\end{centering}
\centering{}\caption{Floquet band structure of twisted bilayer graphene (TBG) below the magic angle. The band-structure shown correspond to  $\theta=0.9^{\circ}$ and $\theta=0.7^{\circ}$ (the first magic angle is at $\theta_{\text{m}}\approx1.1^{\circ}$), and are plotted along the contour shown in Fig.~\ref{fig: illustration}(c). (a),(c) Undriven TBG. At the charge neutrality point, level crossing at the $\Gamma$ point increases the effective bandwidth of the two bands near $E=0$, and no isolated flat bands are observed. (b),(d) Optically driven TBG. The upper and lower bands near $E=0$ become gapped, resulting with nearly flat bands. In panels (b) and (d), we plot the time-averaged density of states $\bar{\rho}_0(E)$,  see Eq.~(\ref{eq:DOS}), for a driving field of frequency   $\hbar\Omega=1.5\,\text{eV}$ and peak electric field of $\mathcal{E}=5.6\,\text{MV/cm}$. Only Floquet states  with significant spectral weight [$A_{\nu}^{0}(\mathbf{k})>0.05$, c.f.  Eq.~(\ref{eq:DOS})] are shown.   \label{fig: band_structure} }
\end{figure}
We model the low energy band-structure of TBG using a continuum model for a single valley and spin \cite {MacDonald2011,Ashvin2019,Kaxiras2019,Fu2018,Senthil2019,Bernevig2019,Cantele2019}. These models, accurately describe the Hamiltonian of TBG with a relatively small twist angle $(\theta\apprle10^{\circ})$ where intervalley processes are strongly suppressed. In the absence of a driving field, our model Hamiltonian is given by
\begin{equation}
H=\left(\begin{array}{cc}
h(\theta/2,\mathbf{r}) & T(\mathbf{r})\\
T^{\dagger}(\mathbf{r}) & h(-\theta/2,\mathbf{r})
\end{array}\right),\label{eq:spatial Hamiltonian}
\end{equation}which acts on the spinor $\Psi(\mathbf{r})=\left(\psi_{1\text{A}},\psi_{1\text{B}},\psi_{2\text{A}},\psi_{2\text{B}}\right)^{T}$. The subscripts $1,2$ denote the top, bottom layer respectively and the $\text{A,B}$ subscripts denote the sub-lattice isospin of a monolayer.  The Hamiltonians of the two rotated monolayers of graphene are denoted by $h(\pm\theta/2,\mathbf{r})$, and feature a nearest-neighbor coupling with a hopping amplitude $\tau$. The operator $T(\mathbf{r})$ denotes the periodic inter-layer moir\'e  potential\[
T(\mathbf{r})=\sum_{n=1}^{3}[w_{0}\sigma_{0}+w_{1}(\sigma_{x}\cos n\phi+\sigma_{y}\sin n\phi)]e^{i(n\phi-\phi-\mathbf{q}_{n}\mathbf{r})}.\]We use the standard Bernal stacking for untwisted layers ($\theta=0$) and $\phi=2\pi/3$. The set of wavenumbers $\mathbf{q}_{1}=k_{\theta}(0,-1)$, $\mathbf{q}_{2,3}=k_{\theta}(\pm\sqrt{3},1)/2$, represents the relative displacements of the Dirac cones between the layers where $k_{\theta}=4\pi\theta/(3\sqrt{3}a)$ is determined by the twist angle $\theta$ and $a=1.42\mathring{\,\text{A}}$. The $2\times2$  Pauli matrices and identity matrix are denoted by  $\boldsymbol{\sigma}$ and $\sigma_{0}$, respectively.  $w_{0}$ denotes the inter-layer coupling between the $AA$ and $BB$ domains and $w_{1}$ denotes the $AB$ and $BA$ inter-layer coupling. Our model uses the exact band-structure of monolayer graphene (generalizing approaches using the $\text{k}\cdot\text{p}$ approximation) for better modeling of the higher energy levels. We use $\tau=2.73\,\text{eV}$, $w_{1}=110\,\text{meV}$, and account for the effects of lattice relaxation by approximating $w_{0}=0.8w_{1}$ \cite{koshino2017,Ashvin2019,Fu2018,Senthil2019}.

We consider a circularly polarized driving light field of frequency $\Omega$, represented by the electric field $\mathbf{\mathcal{E}}(t)=\mathcal{E}(\cos(\Omega t)\hat{x}-\sin(\Omega t)\hat{y})$. We take the light field to be at normal incidence and uniform over the sample. We model the interaction with the driving field using a Peierls substitution for the intra-layer hopping parameters in the Hamiltonian, $\tau\rightarrow\tau\cdot\exp(-ie\mathbf{\mathcal{E}}a/\hbar\Omega)$. In the presence of the time-periodic drive, the solution of the Schrodinger equation can be indexed by the quasi-energies $\varepsilon$, which fall within a single ``Floquet-Brillouin'' zone $-\hbar\Omega/2\leq \varepsilon < \hbar\Omega/2$, and can be written as\begin{equation}
|\psi_{\nu}(t)\rangle=e^{-i\varepsilon_{\nu} t/\hbar}\sum_{m=-\infty}^{\infty}e^{-im\Omega t}|\psi_{\nu}^{(m)}\rangle,\label{eq:Floquet-Spinor}
\end{equation}where the index $\nu$ carries all other quantum numbers of the state.  The set of modes $\sum_m|\psi_{\nu}^{(m)}\rangle$  are the eigenmodes of the Floquet Hamiltonian which we numerically solve in momentum-space, truncating both  the number of Floquet blocks and the infinite representation of the matrix $T(\mathbf{k})$, see Supplementary Material \cite{SM}.
\begin{figure*}[t]
\begin{centering}
\includegraphics[viewport=0bp 0bp 1912bp 677bp,clip,width=17cm]{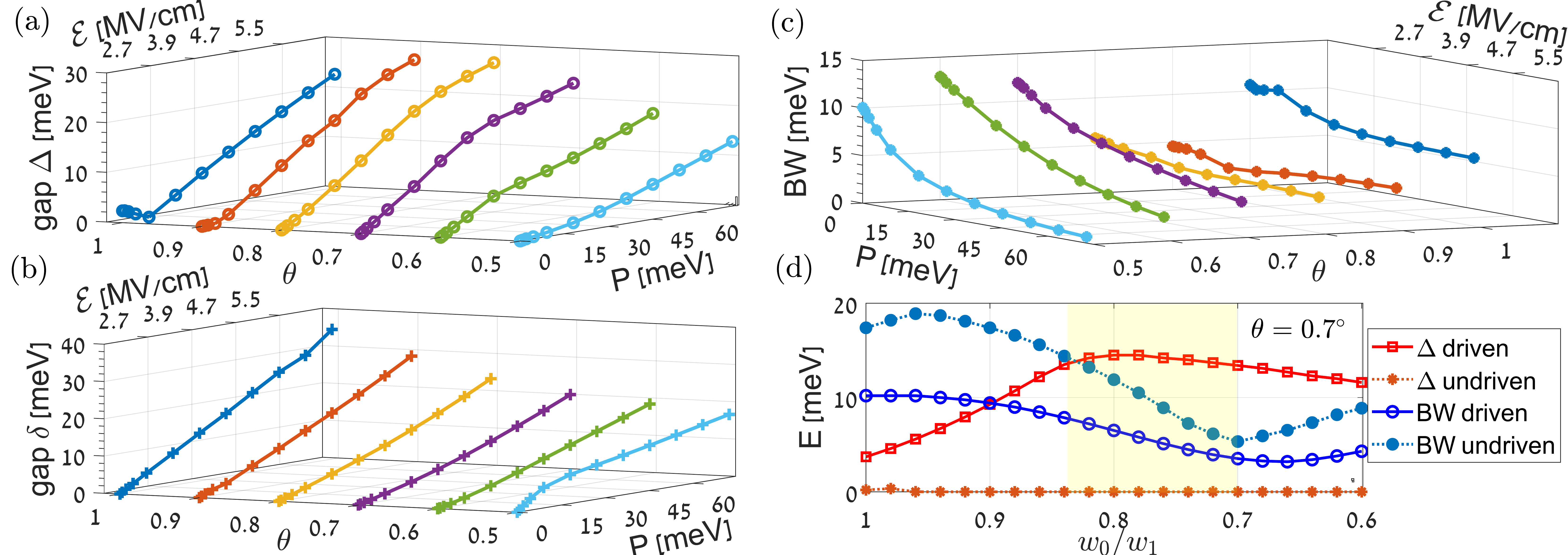}
\par\end{centering}
\centering{}\caption{Characteristics of the upper Floquet-band near the charge neutrality point at $\hbar\Omega=1.5\,\text{eV}$ and angles $\theta\protect\leq1^{\circ}$, below the first magic angle. (a) Increase in the gap $\Delta$ to remote bands with increasing photo-induced gap $P$ (proportional to the optical intensity), which isolates the flat bands. $\mathcal{E}$ denotes the peak electric field of the drive. (b) The gap $\delta$ between the upper and lower bands at the charge neutrality point increases as a function of $P$. (c) Narrowing of the bandwidth $BW$with increasing $P$, at $\theta\protect\leq0.8^{\circ}$. (d) Robustness of the mechanism to variations of lattice relaxation effects, where the yellow shade indicates the range of expected experimental values. The opening of the gap $\Delta$ and reduction in the bandwidth occurs for all physical values of the lattice relaxation parameter $w_0/w_1$.  In all panels, the quantities were calculated from the time-averaged density of states $\bar{\rho}_0(E)$, see Eq.~(\ref{eq:DOS}),  including only Floquet states  with significant spectral weight, $A_{\nu}^{0}(\mathbf{k})>0.05$.  \label{fig:gap-bandwidth-relaxation} }
\end{figure*}

Typical Floquet band-structures of TBG are shown in Fig.~\ref{fig: band_structure}(a)-(d), in the presence and absence of a driving field. The band-structures are plotted along a contour in the first moir\'e Brillouin zone (mBz), which is a hexagon with size $k_{\theta}$, as shown in Fig.~\ref{fig: illustration}(c).  For the undriven case of TBG with twist angle $\theta=0.9^{\circ}$, the lower and upper bands near $E=0$ experience level crossing with other bands at the $\Gamma$ point shown in Fig.~\ref{fig: band_structure}(a). The large bandwidth of the resulting connected group of bands manifests larger kinetic energy of the electrons which hinders the observation of strong correlation effects. Upon driving, an energy gap $\delta$ between the lower and upper bands is opened, as well as an energy gap $\Delta$ isolating these two bands from the rest of the spectrum. These gaps are shown in Fig.~\ref{fig: band_structure}(b) for a drive with $\hbar\Omega=1.5\,\text{eV}$ and peak electric field of $\mathcal{E}=5.6\,\text{MV/cm}$. For $\theta=0.7^{\circ}$, the undriven band-structure exhibits larger bandwidth and multiple level crossings, as shown in Fig.~\ref{fig:band_structure}(c). The drive opens the energy gaps $\delta$ and $\Delta$ and decreases the bandwidth of the lower and upper bands, thus flattening the bands as shown in Fig.~\ref{fig: band_structure}(d). Interestingly, here the drive also flattens the next nearest bands to  CNP.

To quantify the effect of the drive on the band-structure, in Fig.~\ref{fig:gap-bandwidth-relaxation}(a-c) we plot the gaps and bandwidth of the upper Floquet-band at $\hbar \Omega=1.5\text{eV}$ as a function of the twist angle $\theta$ and the quantity  $P=(3\tau ea\mathcal{E})^{2}/(4\hbar^{3}\Omega^3)$,  which gives the photo-induced gap in monolayer graphene (and is proportional to the intensity of the drive at a given frequency) \cite{Floquet-3,Floquet-5}. In the absence of a drive, the energy gap $\delta$ between the lower and upper bands vanishes due to the symmetries of monolayer graphene, and the gap $\Delta$ isolating these two bands vanishes for $\theta\leq0.9^{\circ}$, as shown in Fig.~\ref{fig:gap-bandwidth-relaxation}(a) and Fig.~\ref{fig:gap-bandwidth-relaxation}(b). Upon irradiation, for $\theta\leq0.9^{\circ}$ the  gaps $\Delta$ and $\delta$ obtain nonzero values and increase almost linearly with $P$, thus yielding isolated, narrow bands. In Fig.~\ref{fig:gap-bandwidth-relaxation}(c) we plot the bandwidth of the upper Floquet  band $\varepsilon_{\text{up}}(\mathbf{k})$ given by
$\text{BW}=\text{max}(\varepsilon_{\text{up}}(\mathbf{k}))-\text{min}(\varepsilon_{\text{up}}(\mathbf{k}))$ for $\mathbf{k}$ in the mBz. For $\theta\leq0.8^{\circ}$, the bandwidth of the upper band decreases as the amplitude of the drive is increased. We thus conclude that light irradiation allows for emergence of narrow isolated bands at twist angles smaller than the magic angle.

\begin{figure*}[t]
\begin{centering}
\includegraphics[viewport=0bp 0bp 1476bp 647bp,clip,width=17cm]{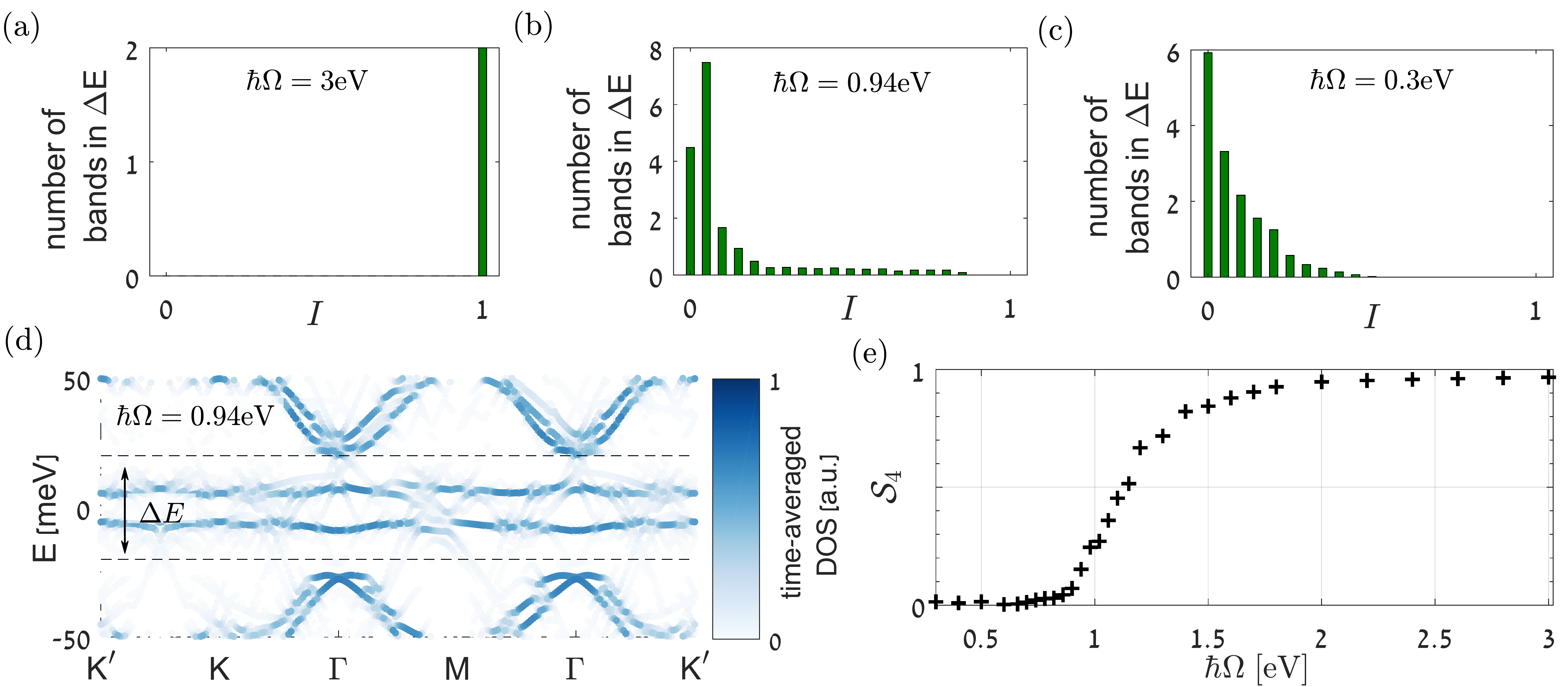}
\par\end{centering}
\centering{}\caption{Mixing with higher energy bands at low driving frequencies.
(a)-(c) Distributions of $I_\nu$, characterizing the time-averaged density of states (DOS) at energies $|E|\protect\leq 20\,\text{meV}$, averaged along the contour $K'\rightarrow K\rightarrow\Gamma\rightarrow M\rightarrow\Gamma\rightarrow K'$ for $\theta=0.9^{\circ}$ and constant photo-induced gap $P=33\,\text{meV}$. (a) For a high drive frequency $\hbar\Omega=3\,\text{eV}$, the time-averaged DOS in the energy interval $|E|\leq20\,\text{meV}$ is sharp, yielding a bimodal distribution peaked at $I=1$ and $I=0$ (not shown).  (b)-(c) For lower frequencies $\hbar\Omega=0.94,0.3\,\text{eV}$ the number of bands with significant spectral weight increases, indicating mixing with higher energy bands. (d) The time-averaged DOS at $\hbar\Omega=0.94\,\text{eV}$, which shows mixing with higher energy bands.  (e) Characterization of the sharpness of the time-averaged DOS as a function of drive-frequency. At drive frequencies in the UV-NIR range $\hbar\Omega\gtrsim1.2\,\text{eV}$, the sharpness parameter $\mathcal{S}_{4}$ approaches unity, indicating that the interaction with the drive is predominately off-resonant. \label{fig:critical-Omeg}}
\end{figure*}

Lattice relaxation effects lead to an expansion of the AB/BA domains at the expense of the AA/BB domains and thus effectively reduce the inter-layer coupling in the latter regions, yielding $w_{0}<w_{1}$. Hence, it is important to verify the robustness of the presented mechanism to changes in the lattice relaxation parameter $w_{0}/w_{1}$. In Fig~\ref{fig:gap-bandwidth-relaxation}(d) and [\onlinecite{SM}], we present such an analysis for $\theta=0.7^{\circ}$ and $\theta=0.9^{\circ}$. For undriven TBG, the gap $\delta$ vanishes at the $K$ point for any value of $w_{0}/w_{1}$, and the gap $\Delta$ vanishes at the $\Gamma$ point within the realistic parameter range of $0.7\leq w_{0}/w_{1}\leq0.85$. The driving field ($P=33\,\text{meV}$) opens both gaps for a wide range of $w_{0}/w_{1}$. The bandwidth is also reduced by the drive in the same range of lattice relaxation parameters. Similar results are obtained for other twist angles below the magic angle. We therefore conclude that the features induced by the drive are robust to variations in the lattice relaxation.

The Floquet bandstructure of TBG in the presence of the drive is a result of both on-resonant and off-resonant processes. Since the number of bands in the reduced mBz for any quasi-momentum is extremely large, it is important to demonstrate that the interaction with the drive, espcially at frequencies below the bandwidth of the material, does not mix the low-energy bands with high-energy bands.  We therefore demonstrate that the presented mechanism is associated with an off-resonant process, and identify the range of frequencies in which the hybridization of the resulting flat bands with high energy bands is suppressed.
To this effect, we define the time-averaged density of states (DOS) defined as \cite{Lindner2019}
\begin{equation}
\bar{\rho}_0(\mathbf{k},E)=\sum_{\nu}\sum_{m} A^{(m)}_\nu(\mathbf{k})\delta(\varepsilon_\nu+m\hbar\Omega-E), \label{eq:DOS}\end{equation}with $A_\nu^{(m)}(\mathbf{k})= |\psi_\nu^{(m)}(\mathbf{k})|^2$. The DOS $\bar{\rho}_0(\mathbf{k},E)$ is the imaginary part of the time-averaged Green's function \cite{Floquet-6,Lindner2019,Gotz2019}.

To quantify the sharpness of the bands near $E=0$,  we integrate $\bar{\rho}_0(\mathbf{k},E)$ in a small interval $\Delta E=40\,\text{meV}$, and compute the total intensity $I(\mathbf{k})=\int_{-\Delta E/2}^{\Delta E/2} dE \bar{\rho}_0(\mathbf{k},E)$ at a given momentum. $I(\mathbf{k})$ consists of discrete contributions from Floquet bands with $2|\varepsilon_\nu|\leq\Delta E$, written as $I(\mathbf{k})=\sum_\nu I_\nu(\mathbf{k})$. In Fig.~\ref{fig:critical-Omeg}(a)-(c) we plot the  distribution of $I_\nu$ averaged along the contour in Fig.~\ref{fig: illustration}(c). We compare the histograms of $I_\nu$ for different drive frequencies at  $\theta=0.9^{\circ}$, maintaining the gaps $\Delta(\Omega)$ and $\delta(\Omega)$ constant throughout the three panels by keeping the photo-induced gap constant, with $P=33\,\text{meV}$.

For drive frequencies in the UV-visible range, the upper and lower flat-bands are sharp. This is indicated by the distribution of $I_\nu$ which has a sharp peak at $I=1$ corresponding to the two flat bands, and no other weights except at $I\approx 0$. Such a distribution is shown in Fig.~\ref{fig:critical-Omeg}(a),  for a driving field in the visible range $(\hbar\Omega=3\,\text{eV})$. For a near-infrared driving field, the spectral weight of the flat bands is reduced while background spectral weight and band-crossings appear, as shown in Fig.~\ref{fig:critical-Omeg}(d), in which we plot  $\bar{\rho}_0(\mathbf{k},E)$ at $\hbar\Omega=0.94\,\text{eV}$. These effects lead to the  broad distribution of $I_\nu$  shown in Fig.~\ref{fig:critical-Omeg}(b), with many Floquet eigenstates corresponding to $0<I_\nu<0.2$ and the two flat Floquet bands corresponding to $I_\nu>0.2$ at most momenta (recall that the distributions $I_\nu$ are averaged along the contour in the mBz). The band-structure in Fig.~\ref{fig:critical-Omeg}(d) can be compared with  the sharply defined bands in Fig.~\ref{fig: band_structure} which shows $\bar{\rho}_0(\mathbf{k},E)$ at $\hbar\Omega=1.5\,\text{eV}$. At even lower driving frequencies, the DOS is dominated by rapidly oscillating Floquet-bands with low spectral weight at the $|E|\leq\Delta E/2$ spectral window, as implied by $I_\nu$ approaching a Poisson distribution, shown in Fig.~\ref{fig:critical-Omeg}(c) for $\hbar\Omega=0.3\,\text{eV}$.

To estimate the range of frequencies for which the DOS at energies $|E|\leq\Delta E/2$ is predominately sharp, we consider the quantity $\mathcal{S}_{n}=\ointctrclockwise \bigl(\sum_\nu \left[I_\nu(\mathbf{k})\right]^n/\sum_\nu I_\nu(\mathbf{k})\bigr)dk$  along the contour in Fig.~\ref{fig: illustration}(c). For $n\geq2$, $\mathcal{S}_n$ approaches unity for a fully sharp DOS, i.e., when the distributions $I_\nu(\mathbf{k})$ are bi-modal and peaked at $I=0$ or $I=1$ at all $\mathbf{k}$'s. Conversely, $\mathcal{S}_n$ with $n\geq2$ becomes vanishingly small when  $I_\nu(\mathbf{k})$ takes the Poisson form peaked at $I=0$, as in Fig~\ref{fig:critical-Omeg}(c). In Fig.~\ref{fig:critical-Omeg}(e) we plot $\mathcal{S}_{4}(\Omega)$, finding that the DOS remains sharp for driving frequencies  $\Omega\gtrsim\Omega^*$ with $\hbar\Omega^*= 1.2\,\text{eV}$, which contains the visible range down to frequencies which are significantly smaller than the graphene bandwidth $\sim17\,\text{eV}$. At drive frequencies in the infrared and below,  $\Omega\lesssim \Omega^*$, the DOS becomes smeared.

To understand the threshold frequency $\Omega^*$, we analyze the mechanism that decreases $\mathcal{S}_4$ for weak interlayer coupling, $\alpha=(w_{0}+w_{1})/(3tk_{\theta})\ll1$ and weak driving $P\ll \hbar \Omega$. When the interlayer coupling is absent ($\alpha=0$)  the Floquet band-structure is that of two driven graphene monolayers as shown in Fig.~\ref{fig: illustration}(b). Note that the Floquet band-structure in Fig.~\ref{fig: illustration}(b) exhibits contours of resonant momenta $\mathbf{k}_R$ which encircle the $K$ points of both monolayers and for which the energy difference between the conduction and valence bands is equal to $\Omega$. At momenta which are deep within this contour (for which~$|\varepsilon_{\nu}|\ll \hbar \Omega$/2), the expansion of the Floquet-states according to Eq.~(\ref{eq:Floquet-Spinor}) is dominated by the $m=0$ component. At the resonant momenta, the original valence and conduction bands with energies $E_{\text{v}}(\mathbf{k})$ and $E_{\text{c}}(\mathbf{k})$ are strongly mixed, and therefore the Floquet states consist of both  $m=0$ and $m=1$ (or $m=-1$) components with similar amplitudes. Moving away from the resonant momenta, the square of the amplitude of the $m=\pm1$ component, which we denote by $\tilde{\mathcal{N}}(\mathbf{k})$, decreases as a Lorentizan with argument $(E_{\text{c}}(\mathbf{k})-E_{\text{v}}(\mathbf{k})-\Omega)$ width $\sqrt{\hbar\Omega P}$ and maximum value of $1/2$.

We now study the effect of non-zero interlayer coupling, $\alpha>0$. The Floquet-Bloch wavefunctions of TBG can be written as\begin{equation}|\psi(\mathbf{k},t)\rangle=\sum_{\nu,n} c_{n,\nu}|\tilde{\psi_\nu}(\mathbf{k_n},t)\rangle
\label{eq: TBG Floquet}
\end{equation}where $|\tilde{\psi}(\mathbf{k},t)\rangle$ are the Floquet states at $\alpha=0$ and $\mathbf{k}_{n}$ is a discrete set of momenta in the extended mBz (see~[\onlinecite{SM}]). Importantly, the coefficients $c_{n,\nu}$ of the flat bands decrease as $|\alpha|^r$ with $r=|\mathbf{k}_{n}|/k_\theta$, since $r$ interlayer tunneling processes are required to connect Floquet states near the $K$ point with Floquet states at $\mathbf{k}_n$. At large drive frequencies, the variation of $I(\mathbf{k})$ along the contour in the mBz is small since the resonant momenta are outside the first mBz, i.e. $|\mathbf{k}_R|\gg k_{\theta}$. We therefore estimate $\mathcal{S}_4\approx |I_0|^3$ where $I_0$ is the contribution to the intensity from one of the Floquet flat bands near the $K$ point. For high frequencies (larger than the graphene bandwidth) the expansion of these states, c.f.~Eq.(\ref{eq:Floquet-Spinor}), is dominated by the $m=0$ component yielding $I_0\to 1$ and $\mathcal{S}_4\to1$. Lowering the frequency below the graphene bandwidth decreases the magnitude of the resonant momenta $|\mathbf{k}_R|$. This leads to an increase in the amplitudes $c_{n,\nu}$ corresponding to the resonant momenta in Eq.~(\ref{eq: TBG Floquet}), which in turn reduces $I_0$.

Using the discussion above, we can estimate the norm of the $m=\pm1$ component of the upper Floquet flat-band as $\mathcal{N}=\sum_{n}\alpha^{2r(\mathbf{k}_n)}\tilde{\mathcal{N}}(\mathbf{k}_n)$. This component does not contribute to $I_0$ and therefore $I_0\approx(1+\mathcal{N})^{-1}$. Taking the largest contribution in the sum $\mathcal{N}$, we arrive at $I_0\approx(1+3r_*\alpha^{2r_*}\tilde{\mathcal{N}}(r_{*}k_\theta))^{-1}$  where $r_*=|\mathbf{k}_R|/k_\theta$, and the factor $3r_{*}$ arises due to the number of $\mathbf{k}_{n}$ momenta with hopping number $r_*$ in the hexagonal grid (see~[\onlinecite{SM}]). For drive frequencies in the visible range, we estimate $|\mathbf{k}_R|\approx\hbar\Omega/(3ta)$ and for $\theta=0.9^{\circ}$ and $P=33\,\text{meV}$ we get $\alpha=0.64$. The condition $\mathcal{S}_{4}\approx|I_0|^{3}\geq0.5$ yields a threshold frequency of $\hbar\Omega^*\approx 1\,\text{eV}$, which can be compared with the numerically calculated  value  $\hbar\Omega^*= 1.2\,\text{eV}$ as shown in Fig.~\ref{fig:critical-Omeg}(e).

For drive frequencies higher than the threshold frequency ($\hbar\Omega\gtrsim1.2\,\text{eV}$), a direct coupling between the bands near CNP and higher levels in the mBZ is suppressed. Thus, the leading order effect of the interaction with the drive can be approximated with the effective static Hamiltonian $H_{\text{eff}}\approx H+[\mathcal{H}_{01},\mathcal{H}_{10}]/\Omega$, describing an off-resonant process which is second order in the drive. The effect of a weak drive $P\ll\hbar\Omega$,  can be  approximated with an addition of a photo-induced Haldane mass term to the Hamiltonian describing each monolayer. Thus in this limit, the Hamiltonians $h_\pm(\mathbf{k})$ describing a monolayer in reciprocal space near the $K(+)$ and $K'(-)$ points acquire an additional term $\eta P\sigma_{z}$, where $\eta=\pm1$ respectively. Therefore, the band structure of the driven system can be described as the result of the inter-layer hybridization between gapped Dirac-cones of the two layers.

\begin{figure}[t]
\begin{centering}
\includegraphics[viewport=10bp 0bp 660bp 300bp,clip,width=8.6cm]{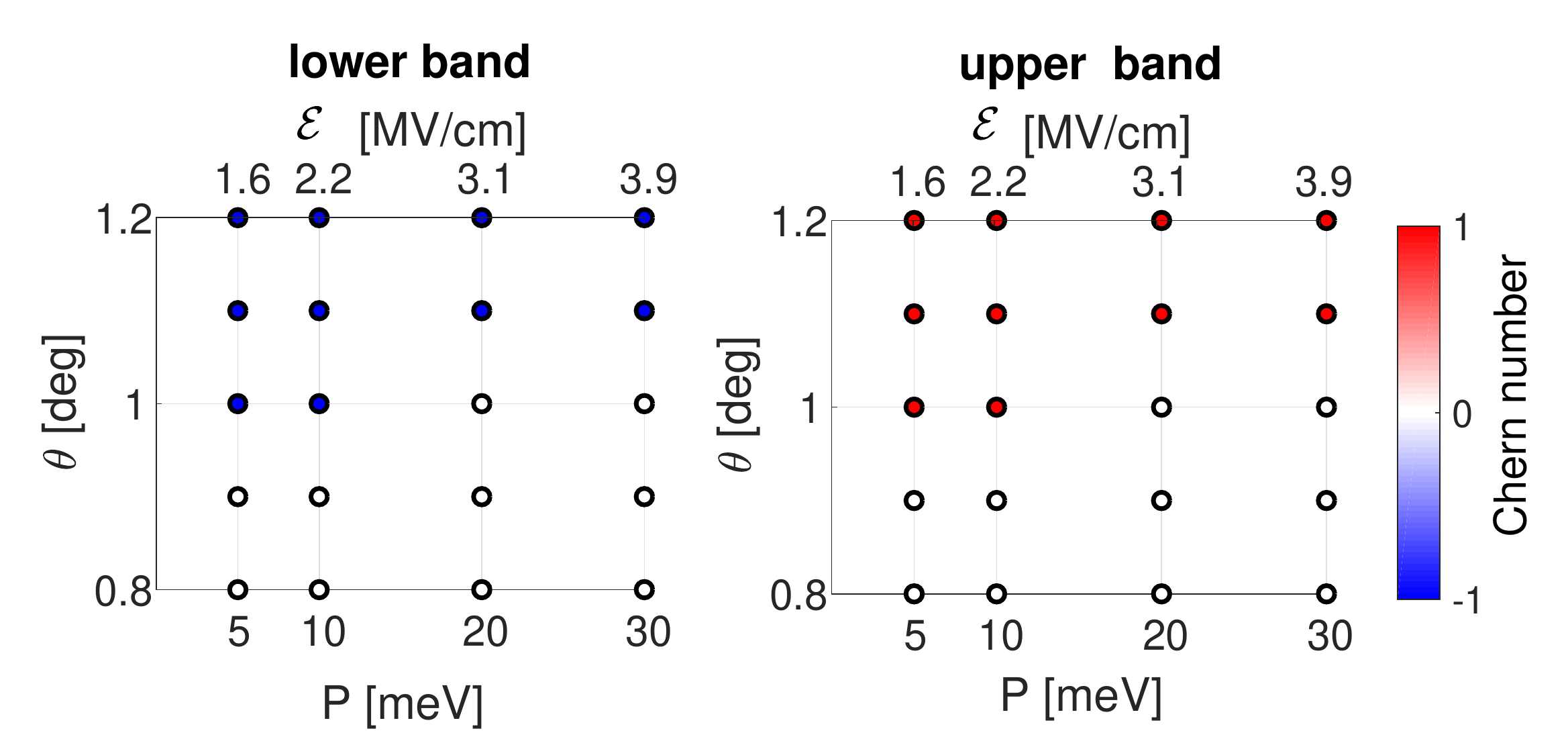}
\par\end{centering}
\centering{}\caption{Chern numbers of flat bands resulting from off-resonant coupling to circularly polarized light.   At twist angles $\theta\protect\geq\theta^*\sim 1^{\circ}$, the lower and upper bands near $E=0$ of TBG with an added Haldane mass term exhibit nonzero Chern numbers, when the magnitude $P$ of the mass term is below a critical value.  The peak electric field $\mathcal{E}$ corresponding to $P$ is given for an optical field of frequency $\hbar\Omega=1.5\,\text{eV}$.  \label{fig:bands_chern}}
\end{figure}

The Haldane mass term breaks time-reversal symmetry and therefore the bands may exhibit non-zero Chern numbers. In undriven TBG, the lower and upper bands cross and thus a Chern number cannot be defined for each of them separately.  Upon driving the system, a nonzero gap $\delta$ between the lower and upper bands  near $E=0$ is opened. We calculate the Chern number of these bands  resulting from the addition of Haldane mass-term near the $K$ point,  [\onlinecite{SM}].  At angles $\theta\geq\theta^*$,  a nonzero Chern number is obtained for weak driving, as shown in Fig.~\ref{fig:bands_chern} for $w_0/w_1=0.8$, which gives $\theta^*\approx1^\circ$.The nonzero Chern number is maintained until a critical drive strength is reached where the bands cross with the nearest remote bands above and below the upper and lower bands. This crossing occurs for $\theta\geq\theta^*$, where the gap $\Delta$ decreases as the drive strength is increased from zero [c.f.~Fig.~\ref{fig:gap-bandwidth-relaxation}(a) at $\theta=1^\circ$]. For larger drive strength than this critical value, the Chern numbers are trivial.

The presented mechanism can be detected experimentally using various techniques. The Floquet band-structure can be observed via spectroscopic methods such as time-resolved angle-resolved photoemission-spectroscopy  \cite{Floquet-9}, and transport properties can be measured via ultrafast electrical measurements \cite{Floquet-10}. The possibility to drive the material with standard pulsed-lasers in the visible-infrared range renders the experimental parameters similar to measurements of single-layer graphene \cite{Sentef2015}.

In conclusion, we show that driving TBG with UV to near-infrared light can lead to appearance of flat bands in the Floquet spectrum. The effect persists at a wide range of twist angles, enabling to engineer flat bands without a need of accurate tuning of the relative twist angle between the graphene layers. Thus Floquet engineering of flat bands may play a particular important role for twisted van-der-Waals heterostructures which typically exhibit long wavelength non-uniformity in the twist angle.

\textit{Note added:} after the initial submission of this manuscript, we became aware of
Ref.~\cite{babak}, which proposes the use of UV-light to obtain topological bands in low angle TBG.

\begin{acknowledgments}
We are grateful to Felix von Oppen for helpful comments and to Shiang Fang for helpful discussion. N.L. and O.K. acknowledge financial support from the  European Research Council (ERC) under the European Union Horizon 2020 Research and Innovation Programme (Grant Agreement No. 639172). We acknowledge support from the IQIM, an NSF physics frontier center funded by Gordon and Betty Moore foundation. We are grateful for support from ARO MURI W911NF-16-1-0361 ``Quantum Materials by Design with Electromagnetic Excitation" sponsored by the U.S. Army. This work was performed in part at Aspen Center for Physics, which is supported by National Science Foundation grant PHY-1607611.
\end{acknowledgments}

\clearpage\newpage{}
\onecolumngrid \appendix
\setcounter{equation}{0}
\setcounter{figure}{0}
\setcounter{table}{0}
\setcounter{page}{1}
\makeatletter
\renewcommand{\theequation}{S\arabic{equation}}
\renewcommand{\thefigure}{S\arabic{figure}}

\part*{Optically induced flat bands in twisted bilayer graphene - Supplementary Material }

\section{Fourier representation of the Hamiltonian}

In this section, we describe the Fourier representation of the Floquet Hamiltonian. The set of modes $|\psi_{\nu}^{(m)}\rangle$ used in the main text are the eigenmodes of the Floquet hamiltonian satisfying the time independent eigenvalue equation
\begin{equation}
\sum_{n=-\infty}^{\infty}\mathcal{H}^{mn}|\psi_{\nu}^{(n)}\rangle=\varepsilon_{\nu}|\psi_{\nu}^{(m)}\rangle.\label{eq:Floquet Eigenvalues}
\end{equation}where $\mathcal{H}^{mn}$ represents a block of the infinite Floquet Hamiltonian in the extended zone, given by
\begin{equation}
\mathcal{H}^{mn}= m\hbar\Omega\delta_{mn}+\frac{\Omega}{2\pi}\int_{0}^{\frac{2\pi}{\Omega}}dte^{-i(m-n)\Omega t}H(t).
\end{equation}We start with the representation of the Hamiltonian of the undriven system.

\subsection{Static Hamiltonian}
We use a convention where the top layer is rotated by an angle $\theta/2$ and the bottom layer is rotated by an angle $-\theta/2$. We generalize the representation in Ref.~\cite{MacDonald2011,thesis_TBG} and represent the Hamiltonian of the first monolayer of graphene rotated by an angle $\theta/2$ in the reciprocal space with \begin{equation}
h_{1}(\theta/2,\mathbf{k}')=-\tau f(\theta/2,\mathbf{k}')\sigma_{1+}+\text{h.c.}
\end{equation}where $\sigma_{1+}=|\mathbf{k}',1A\rangle\langle\mathbf{k}',1B|$  is the Pauli matrix using the iso-spin basis,$1$ denotes the top layer (while $2$ denotes the bottom layer) and $f(\theta/2,\mathbf{k}')=\sum_{i=1}^{3}e^{i\mathbf{k}'\boldsymbol{\delta}_{i}'}$ is the graphene dispersion relation with $\delta_{1}=(0,a)$ and $\delta_{2,3}=a(\mp\sqrt{3},1)/2$. Here and throught thee paper,  primed variable are correspond to the top $(1)$, e.g.  $\mathbf{k'}=R_{z}(\theta/2)\mathbf{k}$ and $\delta_{1}'=R_{z}(\theta/2)\delta_{1}$ whereas $R_{z}(\theta)$ is the rotation matrix around the  $z$ direction which is normal to the plane of the sample. 
Similarly, the Hamiltonian of the second layer is given by\begin{equation} h_{2}(-\theta/2,\mathbf{k}'')=-\tau f(-\theta/2,\mathbf{k}'')\sigma_{2+}+\text{h.c.}\end{equation}where  $\sigma_{2+}=|\mathbf{k}'',2A\rangle\langle\mathbf{k}'',2B|$ and $\mathbf{k}''=R_{z}(-\theta/2)\mathbf{k}$  and $f(-\theta/2,\mathbf{k}'')=\sum_{i=1}^{3}e^{i\mathbf{k}''\boldsymbol{\delta}_{i}''}$. 

The interlayer interaction operator $T(\mathbf{k})$ can be represented by an infinite matrix using the plane wave expansion, with elements $T_{\mathbf{k'},\mathbf{p}''}^{\alpha\beta}\equiv\langle\mathbf{k}',1\alpha|H_{\perp}|\mathbf{p}'',2\beta\rangle$.  The superscripts $\alpha,\beta\in\left\{ A,B\right\} $ denote the iso-spin of the monolayer and the indices $1,2$ indicate the top, bottom layer respectively. Note that $\mathbf{k}'$ and $\mathbf{p}''$ are defined in the repeated reciprocal zone and measured from the center of the Brillouin zone (and not relative to the Dirac point). The interlayer interaction hamiltonian $H_{\perp}$ can be represented  using a tight binding in real space \begin{equation}H_{\perp}=\sum_{ij}t_{\perp ij}\hat{c}^{\dagger}_{1i}\hat{c}_{2j}+\text{h.c.},\end{equation}where $\hat{c}_{1i}$ and $\hat{c}_{2j}$ are the fermionic annihilation operators of the top and bottom layers respectively, the indices $i,j$ denote the real space lattice point,  and $t_{\perp ij}$  denote the  interlayer hopping parameters. We invoke the two-center approximation, assuming that the hopping parameters depend only on the relative distance between the different lattice points  $t_{\perp ij}=t_{\perp}(|\text{R}_{i}-\text{R}_{j}|)$ where $\text{R}_{i}$ denotes the position of the $i^{\text{th}}$ site. Representing $H_{\perp}$ in the reciprocal Fourier space yields the interlayer coupling coefficients \cite{MacDonald2011,thesis_TBG},\begin{align}
T_{\mathbf{k}',\mathbf{p}''}^{\alpha\beta} & =\sum_{\textbf{G}_{1}',\textbf{G}_{2}'}\frac{t_{\perp}(\mathbf{k}'+\textbf{G}_{1}')}{A_{\text{uc}}}e^{i[\textbf{G}_{1}'\delta_{\alpha}'-\textbf{G}_{2}'(\delta_{\beta}'-a)-R_z(-\theta)\textbf{G}_{2}'\textbf{d}]}\delta_{\mathbf{k}'+\textbf{G}_{1}',\mathbf{p}''+R_z(-\theta)\textbf{G}_{2}'}.
\end{align} 
Here we sum over the reciprocal lattice vectors  $\textbf{G}_{1}'$ and $\textbf{G}_{2}'$  of the two layers, $A_{\text{uc}}$ is the unit cell area, $d$  is the spacing between the two layers. The parameters $\boldsymbol{\delta}_{\alpha}'$ and $\boldsymbol{\delta_{\beta}}'$ indicate the relative position of the $A$ and $B$ atoms within a unit cell in the top layer, with  $\boldsymbol{\delta}'_A=\boldsymbol{0}$ and $\boldsymbol{\delta}'_{B}=(0,a)$. The Kronecker delta $\delta_{\mathbf{k}'+\textbf{G}_{1}',\mathbf{p}''+R_z(-\theta)\textbf{G}_{2}'}$ ensures crystal momentum conservation by the tunneling process. 
To simplify the expression of $T_{\mathbf{k}',\mathbf{p}''}^{\alpha\beta}$ further, it is plausible to assume that the interlayer tunneling is slowly varying in space due to the large interlayer separation $d>a$, such that $t_{\perp}(\mathbf{k})$ falls  rapidly to zero at high momenta (cf. \cite{MacDonald2011,thesis_TBG}).  We can then truncate the sum over all reciprocal lattice vectors to a sum over the three smallest reciprocal lattice vectors (which are of the same magntitude) yielding   
\begin{align}
T_{\mathbf{k}',\mathbf{p}''}^{\alpha\beta} & \approx\sum_{i=1}^{3}\delta_{\mathbf{k}'+\mathbf{q}_{i},\mathbf{p}''}T_{i}^{\alpha\beta}.
\label{eq: T_approx}
\end{align}where  the momenta $\mathbf{q}_{i}$ are given in the main text. The $2\times2$ matrices $T_{i}$  in the iso-spin basis which are given in the main text include lattice relaxation effects. Eq.~(\ref{eq: T_approx}) can be considered as a nearest-neighbors hopping hamiltonian in the reciprocal moir\'e lattice.

\begin{figure}[t]
\begin{centering}
\includegraphics[viewport=0bp 0bp 647bp 556bp,clip,width=13cm]{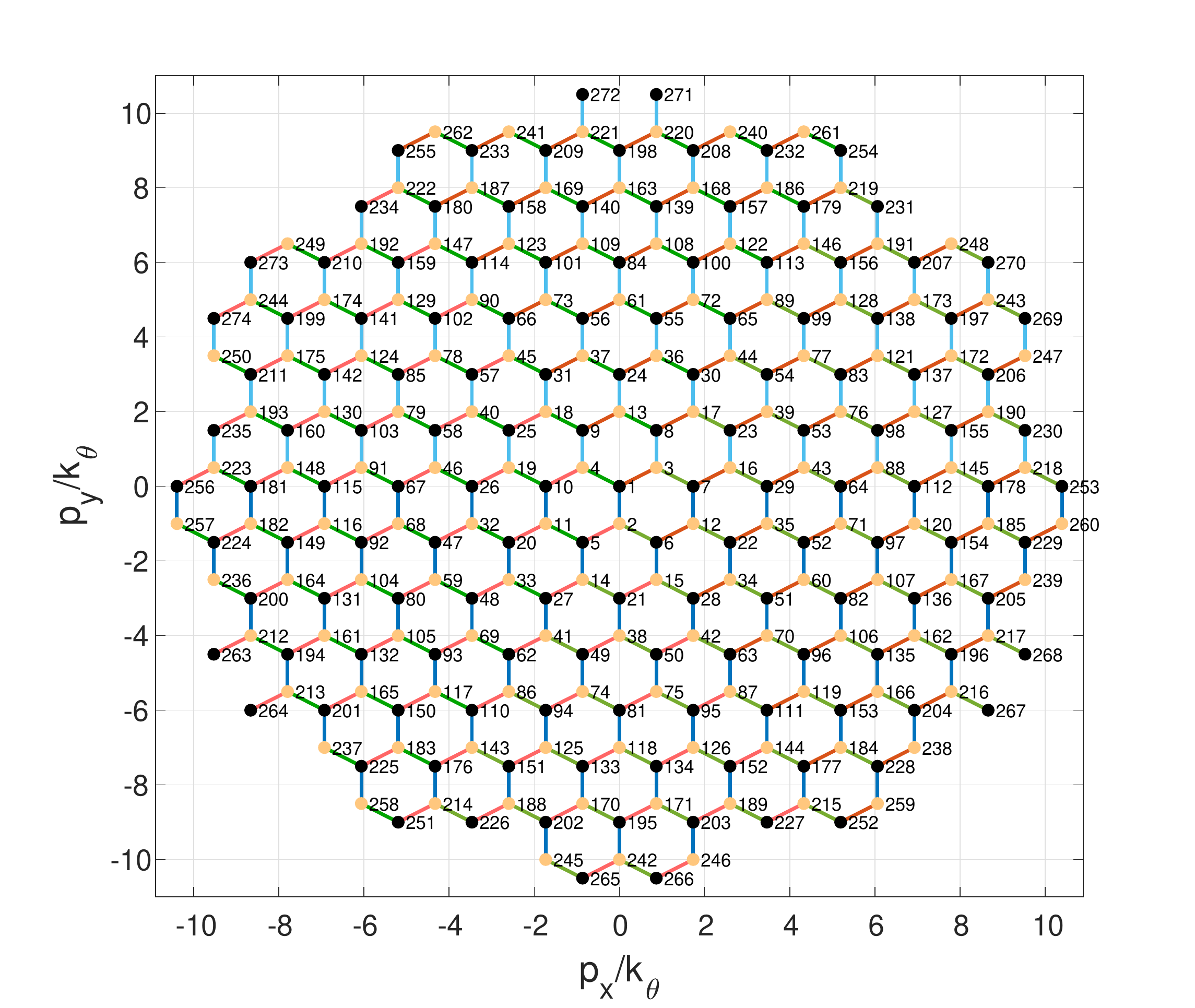}
\par\end{centering}
\centering{}\caption{Reciprocal basis representation of the static Hamiltonian $H(\mathbf{k})$. The nodes represent the monolayer graphene Hamiltonians while the edge describe the interlayer couplings. This representation is truncated with a maximal hopping number $r_{\text{max}}=13$. \label{fig: plane wave expansion} }
\end{figure}

The expansion of $T_{\mathbf{k}',\mathbf{p}''}^{\alpha\beta}$  couples the wavefunction $|\psi_{1}(\mathbf{k}')\rangle$ in the first layer with the wavefunctions of the other layer $|\psi_{2}(\mathbf{p}'')\rangle$ for  $\mathbf{p}''=\mathbf{k}'+\mathbf{q}_{1}$,  $\mathbf{p}''=\mathbf{k}'+\mathbf{q}_{2}$ and  $\mathbf{p}''=\mathbf{k}'+\mathbf{q}_{3}$ . In general, this matrix couples points on a hexagonal lattice in reciprocal space  connected by  reciprocal lattice vectors of the moir\'e lattice. To numerically solve this infinite expansion, we pose a further (numerical) truncation by setting a maximal hopping number $r_{\text{max}}$, where the total number of sites considered is $\tfrac{3}{2}r_{\text{max}}^{2}+\tfrac{3}{2}r_{\text{max}}+1$. As an example, the Hamiltonian for $r_{\text{max}}=1$ is given by
\begin{equation}
H_{k}^{(1)}=\left(\begin{array}{cccc}
h_{1}(\theta/2,\mathbf{k}') & T_{1} & T_{2} & T_{3}\\
T_{1}^{\dagger} & h_{2}(-\theta/2,\mathbf{k}'+\mathbf{q}_{1})\\
T_{2}^{\dagger} &  & h_{2}(-\theta/2,\mathbf{k}'+\mathbf{q}_{2})\\
T_{3}^{\dagger} &  &  & h_{2}(-\theta/2,\mathbf{k}'+\mathbf{q}_{3})
\end{array}\right).\end{equation}A diagrammatic representation of the Hamiltonian in the reciprocal space is shown in Fig.~\ref{fig: plane wave expansion} for $r_{\max}=13$. In this representation, the central node has a momentum $\mathbf{k}$. Each node represents a $2\times2$ Hamiltonian of monolayer graphene, whereas  black nodes represent the matrices $h_{1}(\theta/2,\mathbf{k}'+\mathbf{k}_{n})$ while gold-colored node represent $h_{2}(-\theta/2,\mathbf{k}'+\mathbf{k}_{n})$. Here  $\mathbf{k}_{n}\equiv\mathbf{k}_{lmp}=l\mathbf{g}_{1}+m\mathbf{g}_{2}+p\mathbf{q}_{1}$, denotes the hexagon reciprocal grid in the extended zone. This grid is defined by the moir\'e reciprocal lattice vectors $\mathbf{g}_{1}=\mathbf{q}_1-\mathbf{q}_3,$ and $\mathbf{g}_{2}=\mathbf{q}_2-\mathbf{q}_1$, the integers $l,m$, and the basis index $p=\pm1$. Note that in the main text we denote the different combinations of  $l,m$, and  $p$ with a single index $n$  for brevity.
A nonzero transition matrix connecting layer $1$ to $2$ is represented by an edge, where the transition matrix $T_{1}$ is indicated by a blue edge, a transition matrix $T_{2}$ is indicated by a red edge and a transition matrix $T_{3}$ by a green edge. Note that the coupling between layer $2$ to $1$ is represented in a similar way but with the Hermitian conjugated matrices $T_{1}^{\dagger},T_{2}^{\dagger}$ and $T_{3}^{\dagger}$. The numbers near each point indicate the index of the vector basis in the numerical matrix (not including the iso-spin degree of freedom). It is useful to write the general static Hamiltonian in  reciprocal space with \begin{equation}
H^{(r_{\text{max}})}(\mathbf{k})=T(\mathbf{k})+\sum_{p=0}^{1}\sum_{l,m}|l,m,p\rangle h_{lmp}(\theta/2,\mathbf{k})\langle l,m,p|,\label{eq:reciprocal_abstract_Rep_Hamil}
\end{equation}where the sum is over all integers $l,m$ satisfying $r(l,m,p)=|l|+|m|+|l+m-p|\leq r_{\text{max}}$. The hopping number $r$ counts the number of edges connecting that $\mathbf{k}_{lmp}$ grid point to the  momentum $\mathbf{k}$.   For even values of $r(l,m,p)$ we identify $h_{lmp}=h_{1}(\theta/2,\mathbf{k}'+\mathbf{k}_{lmp})$ while for odd values of $r(l,m,p)$ we have $h_{lmp}=h_{2}(-\theta/2,\mathbf{k}'+\mathbf{k}_{lmp})$.
 In general $r_{\text{max}}$ is determined upon convergence of the result of the low energy (we verify that the error is less than $1\%$ in the low energy spectrum of $|E|\leq50\,\text{meV}$), and we typically use $13\leq r_{\text{max}}\leq22$ in our calculations.

\subsection{Floquet Hamiltonian}
We construct the time dependent driven Hamiltonian by applying the Peierls substitution in reciprocal space $\mathbf{k}\rightarrow\mathbf{k}-e\mathbf{A}(t)/\hbar$, where $\mathbf{A}=\mathbf{\mathcal{E}}/\Omega$ is the vector potential. While the interlayer interaction is independent of the normally incident drive field, the monolayer hamiltonian is transformed via $h_{1}(\theta/2,\mathbf{k}')\rightarrow h_{1}(\theta/2,\mathbf{k}'-e\mathbf{A}'(t)/\hbar)$ for the first layer and $h_{2}(-\theta/2,\mathbf{k}'')\rightarrow h_{2}(-\theta/2,\mathbf{k}''-e\mathbf{A}''(t)/\hbar)$ for the second layer.
We expand the Hamiltonian in a harmonic expansion,
\begin{equation}
h_{1}(\theta/2,\mathbf{k}'-ea\mathbf{A}'(t)/\hbar)=-\tau\sum_{n=-\infty}^{\infty}\left(g_{+}(\theta/2,\mathbf{k}',n)\sigma_{1+}+g_{-}(\theta/2,\mathbf{k}',n)\sigma_{1-}\right)e^{in\Omega t}\label{eq:Floquet Hamiltonian}
\end{equation}where
\begin{equation}
g_{\pm}(\theta/2,\mathbf{k}',n)=i^{n}e^{in(\phi+\theta/2)}\left(e^{\pm i\mathbf{k}'\boldsymbol{\delta}_{1}'}J_{n}(\mp\frac{ea}{\hbar}A)+e^{\pm i\mathbf{k}'\boldsymbol{\delta}_{2}'}e^{-in\psi_{+}}J_{n}(\pm\Gamma_{+})+e^{\pm i\mathbf{k}'\boldsymbol{\delta}_{3}'}e^{-in\psi_{-}}J_{n}(\pm\Gamma_{-})\right),\label{eq:Floquet coefficients}
\end{equation}are the coefficients of the $n^{\text{th}}$ harmonic order for the first layer. The parameters $\Gamma\pm$ and $\psi_\pm$ will be defined below. Similarly for the second layer we use
\begin{equation}
h_{2}(-\theta/2,\mathbf{k}''-ea\mathbf{A}''(t)/\hbar)=-\tau\sum_{n=-\infty}^{\infty}\left(g_{+}(-\theta/2,\mathbf{k}'',n)\sigma_{2+}+g_{-}(-\theta/2,\mathbf{k}'',n)\sigma_{2-}\right)e^{in\Omega t}\label{eq:Floquet Hamiltonian_bot}
\end{equation}where the coefficients of the $n^{\text{th}}$ harmonic order of the second layer are given by
\begin{equation}
g_{\pm}(-\theta/2,\mathbf{k}'',n)=i^{n}e^{in(\phi-\theta/2)}\left(e^{\pm i\mathbf{k}''\boldsymbol{\delta}_{1}''}J_{n}(\mp\frac{ea}{\hbar}A)+e^{\pm i\mathbf{k}''\boldsymbol{\delta}_{2}''}e^{-in\psi_{+}}J_{n}(\pm\Gamma_{+})+e^{\pm i\mathbf{k}''\boldsymbol{\delta}_{3}''}e^{-in\psi_{-}}J_{n}(\pm\Gamma_{-})\right),\label{eq:Floquet coefficients2}
\end{equation}The above relations are derived using the identities \cite{graphene_floquet}\begin{equation}
e^{iz\cos\phi}=\sum_{n=-\infty}^{\infty}i^{n}J_{n}(z)e^{in\phi},\label{eq:identity_floquet_1}
\end{equation}\begin{equation}
e^{-iq\psi}J_{q}(\Gamma)=\sum_{n=-\infty}^{\infty}J_{n+q}(\alpha)J_{n}(\beta)e^{-i\phi n},\label{eq:identity_floquet_2}\end{equation}where $J_{n}(z)$ is the $n^{\text{th}}$ order Bessel function and $\phi$ is the phase retardence between $A_{x}$ and $A_{y}$. The parameters $\Gamma$ and $\psi$ in Eqs.~(\ref{eq:identity_floquet_1})-(\ref{eq:identity_floquet_2}) can be written by two parameters $\alpha$ and $\beta$ as \begin{equation}
\Gamma=\sqrt{\alpha^{2}+\beta^{2}-2\alpha\beta\cos(\phi)},\end{equation}\begin{equation}\psi=\text{atan}\left(\frac{\beta\sin(\phi)}{\alpha-\beta\cos(\phi)}\right).\end{equation}In our case we find that $\Gamma_{\pm}$ and $\psi_{\pm}$ in Eqs.~(\ref{eq:Floquet Hamiltonian})-(\ref{eq:Floquet Hamiltonian_bot})
 are given by\begin{equation}
\Gamma_{\pm}=\frac{a}{2}\sqrt{A_y^{2}+3A_x^{2}\pm 2\sqrt{3}A_x A_y cos(\phi)},\end{equation}and
\begin{equation}\psi_{\pm}=\text{atan}\left(\frac{\sqrt{3}A_x\sin(\phi)}{\pm A_y+\sqrt{3} A_x\cos(\phi)}\right).\end{equation}For the special case of circularly polarized light considered here we use $\phi=\pi/2$ and obtain the simple relations $\Gamma_{\pm}=eaA/\hbar$ and $\psi_{\pm}=\pm\pi/3$. We can use the expansion in Eq.~(\ref{eq:Floquet Hamiltonian}), with a truncated sum on integers from $-N_F$ to $N_F$, in order to construct the time independent Floquet Hamiltonian $\mathcal{H}_{m n}$ with $-N_{\text{F}}\leq m\leq N_{\text{F}}$, yielding\begin{equation}
\mathcal{H}_{m,m+n}(\mathbf{k})=\bigl(m\Omega\sigma_{0}+T(\mathbf{k})\bigr)\delta_{mn}+\sum_{p'=0}^{1}\sum_{l',m'}|l',m',p'\rangle \tilde{h}_{l'm'p'}(\theta/2,\mathbf{k},n)\langle l',m',p'|.
\end{equation}where for even values of $r(l',m',p')$ we define \begin{equation}\tilde{h}_{l'm'p'}(\theta/2,\mathbf{k}',n)=-\tau \left(g_{+}(\theta/2,\mathbf{k}'+\mathbf{k}_{l'm'p'},n)\sigma_{1+}+g_{-}(\theta/2,\mathbf{k}'+\mathbf{k}_{l'm'p'},n)\sigma_{1-}\right),\end{equation}and for odd values of $r(l',m',p')$ we define \begin{equation}\tilde{h}_{l'm'p'}(\theta/2,\mathbf{k}',n)=-\tau \left(g_{+}(-\theta/2,\mathbf{k}'+\mathbf{k}_{l'm'p'},n)\sigma_{2+}+g_{-}(-\theta/2,\mathbf{k}'+\mathbf{k}_{l'm'p'},n)\sigma_{2-}\right).\end{equation}In practice, we observe convergence of numerical calculations when setting a cutoff of $N_{\text{F}}=2$ or $3$. The dimension of the Floquet matrix is then $(2N_{\text{F}}+1)\cdot(3r_{\text{max}}^{2}+3r_{\text{max}}+2)$.

The functions $g_{\pm}(\pm\theta/2,\mathbf{k},n)$ obtains a simple form in the low power regime $eaA\ll\hbar$. In this regime the zeroth order describes the undriven Hamiltonian with $g_{+}(\pm\theta/2,\mathbf{k},0)=f(\pm\theta/2,\mathbf{k})$ and $g_{-}(\pm\theta/2,\mathbf{k},0)=f^{*}(\pm\theta/2,\mathbf{k})$. Furthermore near the $K$ point in the top layer we find that $g_{\pm}(\theta/2,\mathbf{k},-1)=0$ and $g_{\pm}(\theta/2,\mathbf{k},1)\approx 3eaAe^{i\theta/2}/(2\hbar)$, and near the $K$ Dirac point of the bottom layer we have $g_{\pm}(-\theta/2,\mathbf{k},-1)=0$ and $g_{\pm}(\theta/2,\mathbf{k},1)\approx 3eaAe^{-i\theta/2}/(2\hbar)$.  We also find that the contribution of terms with higher orders of $n$ are smaller by powers of $(eaA/\hbar)^{n}$. Importantly, for $\theta\ll1$ we find that the effective Hamiltonian in leading powers of the small parameter $[(3\tau eaA)/(2\hbar^2\Omega)]$ at these conditions is given by\begin{equation} H_{\text{eff}}\approx H+[\mathcal{H}_{01},\mathcal{H}_{10}]/\Omega\end{equation}
\clearpage
\section{Lattice relaxation}

\begin{figure}[h]
\begin{centering}
\includegraphics[viewport=0bp 0bp 957bp 315bp,clip,width=16cm]{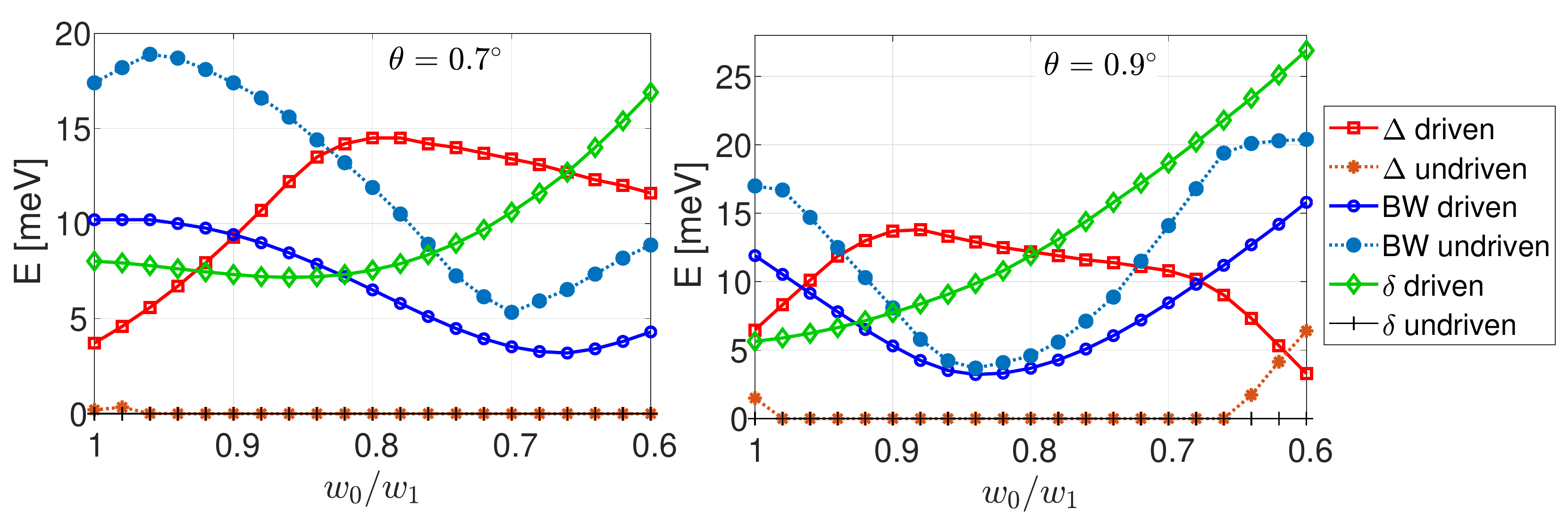}
\par\end{centering}
\centering{}\caption{The increased gaps $\Delta,\delta$  and low bandwidth of driven TBG are robust with variations of lattice relaxation effects. Calculation for $\theta=0.7^{\circ}$ (left) and $\theta=0.9^{\circ}$  (right) using  $\hbar\Omega=1.5\,\text{eV}$ and $\mathcal{E}=4\,\text{MV/cm}$.\label{fig:lattice_relaxation2} }
\end{figure}

In Fig.~\ref{fig:lattice_relaxation2}, we calculate the dependence of the bandwidth and the gaps $\Delta$,  $\delta$ on the parameter $w_{0}/w_{1}$ for $\theta=0.7^{\circ}, 0.9^{\circ}$ using  $\hbar\Omega=1.5\,\text{eV}$ and $\mathcal{E}=4\,\text{MV/cm}$ ($P=33\,\text{meV}$). These results demonstrate that the effect of the driving weakly depends on the exact lattice relaxation parameter.

\section{Topology of the flat bands}
\setcounter{equation}{23}
We calculate the Chern number of the upper an lower bands  near $E=0$ resulting from the addition of Haldane mass-term near the $K$ point, by following the procedure presented in Refs.~\cite{Senthil_Chern2019_SM,Japan_Berry_SM}. We discretize the reciprocal space using a rectangular grid $\mathbf{k}=k_{\theta}(m\hat{x}+n\hat{y})/N_{k}$ for integer $-N_{k}\leq m,n\leq N_{k}$ and typically use $N_{k}=200$. We consider only the grid points within the first mBz hexagon. Our goal is to diagonalize the static Hamiltonian in Eq.~(\ref{eq:reciprocal_abstract_Rep_Hamil}) including a Haldane photo-induced mass term of size $P$.

\begin{figure}[t]
\begin{centering}
\includegraphics[viewport=0bp 0bp 730bp 360bp,clip,width=15cm]{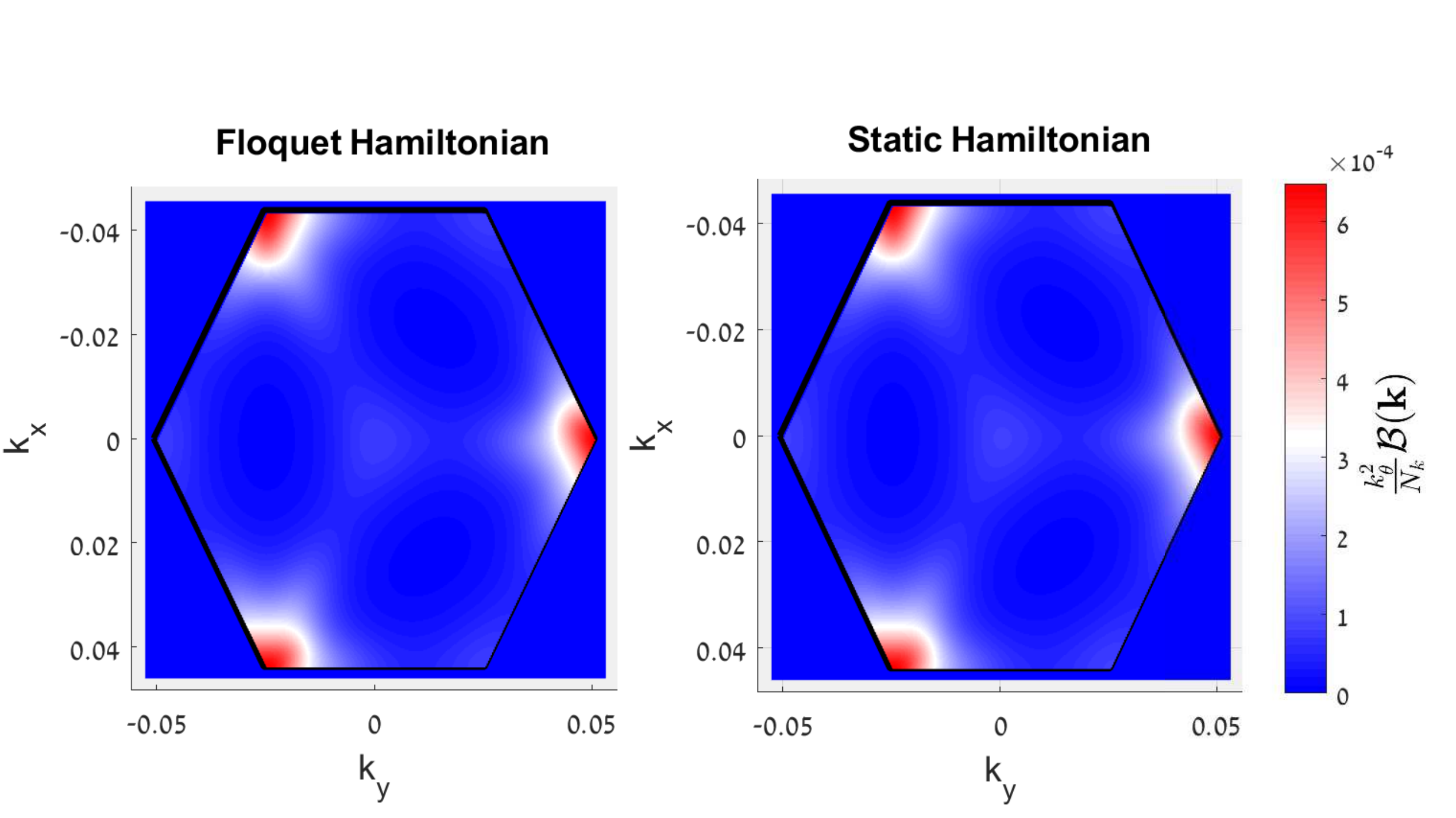}
\par\end{centering}
\centering{}\caption{Berry curvature of the upper band in the mBz for $\theta=1.2^{\circ}$
and $P=10\,\text{meV}$. The effective static Hamiltonian and the
full Floquet analysis yield the same result.\label{fig:Berry_curv} }
\end{figure}

Recall that the static Hamiltonian Eq.~(\ref{eq:reciprocal_abstract_Rep_Hamil}) is valley and spin degenerate. The two valley degenerate bands arise from Bloch wavefunctions which are superpositions of momenta near the $K$ or $K'$ points of both layers. Here, we calculate the Bloch wavefunctions in the mBz corresponding to the bands whose momenta $\mathbf{k}$  are always near the $\mathbf{K}$ points of both layers, and far from $\mathbf{K'}$ points of both layers, since $r_{\text{max}} k_\theta\ll |\mathbf{K}-\mathbf{K'}|$).  Therefore, to take the Haldane mass term into account, it is sufficient to add to the Hamiltonian $h(k)$ appearing in  Eq.~(\ref{eq:reciprocal_abstract_Rep_Hamil}) a mass term $P\sigma_z$, which represents the Haldane mass term correctly near the $K$ point.  

We diagonalize the Hamiltonian in Eq.~(\ref{eq:reciprocal_abstract_Rep_Hamil}) including the term $P\sigma_z\otimes\mathcal{\sigma_0}$ and find the wavefunctions $|\psi_{l}(\mathbf{k})\rangle$ of the lower and upper bands denoted with $l\in\left\{ v,c\right\} $. We then calculate the local Berry curvature with \begin{equation}
\mathcal{B}_{l}(\mathbf{k})=\left(\frac{N_{k}}{k_{\theta}}\right)^{2}\arg\left(U_{l,x}(\mathbf{k})U_{l,y}(\mathbf{k}+\hat{x}\tfrac{k_{\theta}}{N_{k}})U_{l,x}^{*}(\mathbf{k}+\hat{y}\tfrac{k_{\theta}}{N_{k}})U_{l,y}^{*}(\mathbf{k})\right),
\end{equation}where the potentials $U_{l,x}(\mathbf{k})$ and $U_{l,y}(\mathbf{k})$ are given by\begin{equation}
U_{x}^{v,c}(\mathbf{k})=\langle\psi_{v,c}(\mathbf{k}+\hat{x}k_{\theta}/N_{k})|\psi_{v,c}(\mathbf{k})\rangle,
\end{equation}and\begin{equation}
U_{y}^{v,c}(\mathbf{k})=\langle\psi_{v,c}(\mathbf{k}+\hat{y}k_{\theta}/N_{k})|\psi_{v,c}(\mathbf{k})\rangle.
\end{equation}

We calculate the Chern number of the lower and upper bands in Fig.~5 in the main texy by summing the local Berry curvature over the mBz \begin{equation}
C_{l}=\frac{1}{2\pi}\left(\frac{k_{\theta}}{N_{k}}\right)^{2}\sum_{\mathbf{k}}\mathcal{B}_{l}(\mathbf{k}).
\end{equation}
An example of the Berry curvature of the upper band of TBG with $\theta=1.2^{\circ}$ driven by $P=10\,\text{meV}$ is shown in Fig.~\ref{fig:Berry_curv} for the static Hamiltonian. 

At relatively high drive frequencies, as shown in the main text, the Floquet bands closely resemble the bands of the static Hamiltonian, as mixing and level crossings are suppressed. Therefore we expect that the Berry curvature of the Floquet bands should closely resemble the Berry curvature of the static calculation with the Haldane mass term, up to isolated level crossings which effect the Berry curvature in very narrow regions in the reciprocal space. To demonstrate this we calculate the Berry curvature of the Floquet bands with the full time dependent Hamiltonian with $\hbar\Omega=3\,\text{eV}$ , in Fig.~\ref{fig:Berry_curv}. To perform this calculation we chose the wavefunctions of the upper and lower bands as the two wavefunctions whose spectral weight  $A_{\nu}^{0}(\mathbf{k})$ in the energy interval $|E|<\Delta E$ is maximal, c.f.~Eq.~(3) in the main text (at high drive frequencies, these bands have $A_{\nu}^{0}(\mathbf{k})\approx 1 $).  This example demonstrates that indeed the Berry curvature of the static Hamiltonian captures the one of the Floquet bands at relatively high drive frequencies. 


\begin{thebibliography}{45}
\bibitem{geim_VDW_review}A. K. Geim, and I. V. Grigorieva, \textit{Van der Waals heterostructures}, Nature 499, 419-425 (2013).

\bibitem{Novoselov2016}  K. S. Novoselov, A. Mishchenko, A. Carvalho, and A. H. C. Neto, \textit{2D materials and van der Waals heterostructures}, Science 353, aac9439 (2016).

\bibitem{VDW_review_2016} Y. Liu, N. O. Weiss, X. Duan, H.-C. Cheng, Y. Huang, and X. Duan, \textit{Van der Waals heterostructures and devices},  Nat. Rev. Mater. 1, 16042 (2016).

\bibitem{MacDonald_2019} A. H. MacDonald, \textit{Trend: Bilayer Graphene's Wicked, Twisted Road}, Physics 12, 12 (2019).

\bibitem{Neto_2007} J. M. B. Lopes dos Santos, N. M. R. Peres, and A. H. Castro Neto, \textit{Graphene Bilayer with a Twist: Electronic Structure}, Phys. Rev. Lett. 99, 256802 (2007).

\bibitem{Dean2018} R. Ribeiro-Palau, C. Zhang, K. Watanabe, T. Taniguchi, J. Hone, and C. R. Dean, \textit{Twistable electronics with dynamically rotatable heterostructures}, Science 361, 690 (2018).

\bibitem{Kaxiras2017} S. Carr, D. Massatt, S. Fang, P. Cazeaux, M. Luskin, and E. Kaxiras, \textit{Twistronics: Manipulating the electronic properties of two-dimensional layered structures through their twist angle}, Phys. Rev. B 95, 075420 (2017).

\bibitem{Neto_2012} J. M. B. Lopes dos Santos, N. M. R. Peres, and A. H. Castro Neto, \textit{Continuum model of the twisted graphene bilayer}, Phys. Rev. B 86, 155449 (2012).

\bibitem{MacDonald2011} R. Bistritzer and A. H. MacDonald, \textit{Moire bands in twisted double-layer graphene}, Proc. Natl. Acad. Sci. U.S.A. 108, 12233 (2011).

\bibitem{Ashvin2019} G. Tarnopolsky, A. J. Kruchkov, and A. Vishwanath, \textit{Origin of Magic Angles in Twisted Bilayer Graphene}, Phys. Rev. Lett. 122, 106405 (2019).

\bibitem{Song2019} L. Shi, J. Ma and J. C.W. Song, \textit{Gate-tunable flat bands in van der Waals patterned dielectric superlattices}, arXiv:1904.07877 (2019).

\bibitem{PJH_superconductivity} Y. Cao, V. Fatemi, S. Fang, K. Watanabe, T. Taniguchi, E. Kaxiras, and P. Jarillo-Herrero, \textit{Unconventional superconductivity in magic-angle graphene superlattices}, Nature 556, 43 (2018).

\bibitem{PJH_correlated_ins} Y. Cao, V. Fatemi, A. Demir, S. Fang, S. L. Tomarken, J. Y. Luo, J. D. Sanchez-Yamagishi, K. Watanabe, T. Taniguchi, E. Kaxiras, R. C. Ashoori, P. Jarillo-Herrero, \textit{Correlated insulator behaviour at half-filling in magic-angle graphene superlattices}, Nature (London) 556, 80 (2018).

\bibitem{Macdonald2017} K Kim, A. DaSilva, S. Huang, B. Fallahazad, S. Larentis, T. Taniguchi, K. Watanabe, B. J. LeRoy, A. H. MacDonald, and E. Tutuc, \textit{Tunable moire bands and strong correlations in small-twist-angle bilayer graphene}, Proc. Natl. Acad. Sci. U.S.A. 114, 3364 (2017).

\bibitem{Dean_2019} M. Yankowitz, S. Chen, H. Polshyn, Y. Zhang, K. Watanabe, T. Taniguchi, D. Graf, A. F. Young, and C. R. Dean, \textit{Tuning superconductivity in twisted bilayer graphene}, Science 363, 1059 (2019).

\bibitem{DGG_ferro_2018} A. L. Sharpe, E. J. Fox, A. W. Barnard, J. Finney, K. Watanabe, T. Taniguchi, M. A. Kastner, and D. Goldhaber-Gordon, \textit{Emergent ferromagnetism near three-quarters filling in twisted bilayer graphene}, Science 365, 605-608 (2019).

\bibitem{Bernevig_phonons} B. Lian, Z. Wang, and B. A. Bernevig, \textit{Twisted Bilayer Graphene: A Phonon-Driven Superconductor}, Phys. Rev. Lett. 122, 257002 (2019).

\bibitem{koshino2017} N. N. T. Nam and M. Koshino, \textit{Lattice relaxation and energy band modulation in twisted bilayer graphene}, Phys. Rev. B 96, 075311 (2017).

\bibitem{Floquet-1} G. Floquet, \textit{Sur les equations differentielles lineaires a coefficients periodiques}, Ann. de l'Ecole Norm. Suppl. 12, 47 (1883).

\bibitem{Floquet-2} W. Yao, A. H. MacDonald, and Q. Niu, \textit{Optical Control of Topological Quantum Transport in Semiconductors}, Phys. Rev. Lett. 99, 047401 (2007).

\bibitem{Floquet-3} T. Kitagawa, E. Berg, M. Rudner, and E. Demler, \textit{Topological characterization of periodically driven quantum systems}, Phys. Rev. B 82, 235114 (2010).

\bibitem{Floquet-4} N. H. Lindner, G. Refael, and V. Galitski, \textit{Floquet Topological Insulator in Semiconductor Quantum Wells}, Nature Phys. 7, 490 (2011).

\bibitem{Floquet-5} T. Oka and H. Aoki, \textit{Photovoltaic Hall effect in graphene}, Phys. Rev. B 79, 081406 (2009).

\bibitem{Floquet-6} G. Usaj, P. M. Perez-Piskunow, L. E. F. Foa Torres, and C. A. Balseiro, \textit{Irradiated graphene as a tunable Floquet topological insulator}, Phys. Rev. B 90, 115423 (2014).

\bibitem{Floquet-7} T. Kitagawa, T. Oka, A. Brataas, L. Fu, and E. Demler, \textit{Transport properties of nonequilibrium systems under the application of light: Photoinduced quantum Hall insulators without Landau levels}, Phys. Rev. B 84, 235108 (2011).

\bibitem{Floquet-8} N. H. Lindner, D. L. Bergman, G. Refeal, and V. Galitski, \textit{Topological Floquet spectrum in three dimensions via a two-photon resonance}, Phys. Rev. B 87, 235131 (2013).

\bibitem{Floquet-9} Y. H. Wang, H. Steinberg, P. Jarillo-Herrero, and N. Gedik, \textit{Observation of Floquet-Bloch states on the surface of a topological insulator}, Science 342, 453 (2013).

\bibitem{Floquet-10} J. W. McIver, B. Schulte, F.-U. Stein, T. Matsuyama, G. Jotzu, G. Meier, and A. Cavalleri, \textit{Light-induced anomalous Hall effect in graphene}, Nat. Phys. (2019). doi:10.1038/s41567-019-0698-y.

\bibitem{Floquet-11} P. M. Perez-Piskunow, G, Usaj, C. A. Balseiro, and L. E. F. Foa Torres, \textit{Floquet chiral edge states in graphene}, Phys. Rev. B 89, 121401(R) (2014).

\bibitem{Moore-2017} W. Berdanier, M. Kolodrubetz, R. Vasseur, and J. E. Moore, \textit{Floquet Dynamics of Boundary-Driven Systems at Criticality}, Phys. Rev. Lett. 118, 260602 (2017).

\bibitem{Fertig2011} Z. Gu, H. A. Fertig, D. P. Arovas, and A. Auerbach, \textit{Floquet Spectrum and Transport through an Irradiated Graphene Ribbon}, Phys. Rev. Lett. 107, 216601 (2011).

\bibitem{Sentef2015} M. A. Sentef, M. Claassen, A. F. Kemper, B. Moritz, T. Oka, J. K. Freericks, and T. P. Devereaux, \textit{Theory of pump-probe photoemission in graphene: Ultrafast tuning of Floquet bands and local pseudospin textures}, Nature Comm. 6, 7047 (2015).

\bibitem{Klinovaja2016} J. Klinovaja, P. Stano, and D. Loss, \textit{Topological Floquet Phases in Driven Coupled Rashba Nanowires}, Phys. Rev. Lett. 116, 176401 (2016).

\bibitem{Liu2018} J. Liu, K. Hejazi, and L. Balents, \textit{Floquet Engineering of Multiorbital Mott Insulators: Applications to Orthorhombic Titanates}, Phys. Rev. Lett. 121, 107201 (2018).

\bibitem{Lindner2019} M. S. Rudner, N. H. Lindner, \textit{Floquet topological insulators: from band structure engineering to novel non-equilibrium quantum phenomena}, arXiv:1909.02008 (2019).

\bibitem{Floquet_graphene} G. E. Topp, G. Jotzu, J. W. McIver, L. Xian, A. Rubio, and M. A. Sentef, \textit{Topological Floquet engineering of twisted bilayer graphene}, Phys. Rev. Research 1, 023031 (2019).

\bibitem{dynamic_localization}D. H. Dunlap and V. M. Kenkre, \textit{Dynamic localization of a charged particle moving under the influence of an electric field}, Phys. Rev. B 34, 3625  (1986).

\bibitem{Kaxiras2019} S. Carr, S. Fang, Z. Zhu, and E. Kaxiras, \textit{Exact continuum model for low-energy electronic states of twisted bilayer graphene}, Phys. Rev. Research 1, 013001 (2019).

\bibitem{Fu2018} M. Koshino, N. F. Q. Yuan, T. Koretsune, M. Ochi, K. Kuroki, and L. Fu, \textit{Maximally Localized Wannier Orbitals and the Extended Hubbard Model for Twisted Bilayer Graphene}, Phys. Rev. X 8, 031087 (2018).

\bibitem{Senthil2019} Y. Zhang, D. Mao, and T. Senthil, \textit{Twisted Bilayer Graphene Aligned with Hexagonal Boron Nitride: Anomalous Hall Effect and a Lattice Model}, arXiv preprint, arXiv:1901.08209 (2019).

\bibitem{Bernevig2019} Z. Song, Z. Wang, W. Shi, G. Li, C. Fang, and B. A. Bernevig, \textit{All Magic Angles in Twisted Bilayer Graphene are Topological}, Phys. Rev. Lett. 123, 036401 (2019).

\bibitem{Cantele2019} P. Lucignano, D. Alf\'e, V. Cataudella, D. Ninno, and G. Cantele, \textit{Crucial role of atomic corrugation on the flat bands and energy gaps of twisted bilayer graphene at the magic angle}, Phys. Rev. B 99, 195419 (2019).

\bibitem{SM} See Supplementary Material for: i) Fourier representation of the Floquet Hamiltonian, ii) calculation of the lattice relaxtion effects at different twist angles and, iii) calculation of the Chern numbers and the Berry curvature.

\bibitem{Senthil_Chern2019} Y. Zhang, D. Mao, Y. Cao, P. Jarillo-Herrero, and T. Senthil, \textit{Nearly flat Chern bands in moire superlattices}, Phys. Rev. B 99, 075127 (2019).

\bibitem{Japan_Berry} T. Fukui, Y. Hatsugai, and H. Suzuki, \textit{Chern Numbers in Discretized Brillouin Zone: Efficient Method of Computing (Spin) Hall Conductances}, J. Phys. Soc. Jpn. 74, 1674 (2005).

\bibitem{Gotz2019} G. S. Uhrig, M. H. Kalthoff, and J. K. Freericks, \textit{Positivity of the Spectral Densities of Retarded Floquet Green Functions}, Phys. Rev. Lett. 122, 130604 (2019).

\bibitem{babak}  Y. Li, H. A. Fertig, B. Seradjeh, \textit{Floquet-Engineered Topological Flat Bands in Irradiated Twisted Bilayer Graphene}, arXiv:1910.04711 (2019).



\bibitem{thesis_TBG} G. Catarina, B. Amorim, E. V. Castro, J. M. V. P. Lopes and N. M. R. Peres. Handbook of Graphene: Volume 3, Edited by Mei Zhang (John Wiley \& Sons, New Jersey, 2019), Chap. 6, pp. 177-230.

\bibitem{graphene_floquet} P. Delplace,  \'A. G\'omez-Le\'on and G. Platero, Phys. Rev. B  88, 245422 (2013).

\bibitem{Senthil_Chern2019_SM}Y. Zhang, D. Mao, Y. Cao, P. Jarillo-Herrero, and T. Senthil, Phys. Rev. B 99, 075127 (2019).

\bibitem{Japan_Berry_SM} T. Fukui, Y. Hatsugai, and H. Suzuki, J. Phys. Soc. Jpn. 74, 1674 (2005).


\end{thebibliography}
\end{document}